\documentclass[dvips]{article}

\usepackage{icrctc07}

\title{Photon spectra from final stages of a primordial black hole evaporation in different theoretical models}

\shorttitle{Photon spectra from final stages of a primordial black
hole evaporation}

\authors{Edgar Bugaev, Peter Klimai, Valery Petkov}

\shortauthors{Bugaev et al.}

\afiliations{Institute for Nuclear Research, Moscow, Russia}

\email{bugaev@pcbai10.inr.ruhep.ru}

\abstract{Possibilities of an experimental search for gamma-ray
bursts from primordial black hole (PBH) evaporations in space are
reconsidered. It is argued that the corresponding constraints
which can be obtained in experiments with cosmic ray detectors
strongly depend on theoretical approach used for a description of
the PBH evaporation process. Predictions of several theoretical
models for gamma-ray spectra from final stages of PBH life
(integrated over time) are given.}

\begin{document}
\maketitle

\section{Introduction}

Direct searches for the bursts of gamma rays from the evaporations
of PBHs have been carried out in several works during last decade
\cite{Alexandreas:1993zx, Amenomori, Connaughton:1998ac, Funk,
Linton:2006yu}. PBHs are a unique probe of general relativity,
cosmology and quantum gravity \cite{Carr:2003bj}. They could have
been formed in the early universe, e.g., from direct gravitational
collapse of primordial density perturbations. It was shown in
numerous works that PBHs can be produced due to specific features
of the inflationary potential or due to perturbation amplification
during preheating phase. For direct searches of PBHs their spatial
clustering properties are very essential. It was shown recently
\cite{Chisholm:2005vm} that the local clustering enhancement of
PBH number density in our galaxy can be very large, up to $\sim
10^{22}$, i.e., many orders of magnitude larger than the factors
$\sim 10^7$ predicted earlier in some works. Due to this, the
limits from direct searches of PBHs can be much stronger than
indirect limits from measurements of diffuse gamma ray
extragalactic background.

\section{PBH evaporation models}

\subsection{Model of MacGibbon and Webber}

In first calculations of photon spectra from BH evaporations it
had been shown that this spectrum is far from being thermal simply
because emitted elementary particles such as quarks fragment into
hadrons, photons, neutrinos, etc, and just this fragmentation and
decay of unstable hadrons produces the final photon spectrum
\cite{MacGibbon:1990zk}. It was assumed that the radiation
evaporated from the BH interacts too weakly and all emitted quarks
propagate freely and fragment independently of each other.

The resulting time-integrated spectrum (for the case $T_H \gg
m_\pi$) can be parametrized in the simple form
\begin{eqnarray}
 \frac{dN}{dE_{\gamma}} \approx 3 \times 10^{20} \left\{ \begin{array}
{l}
  \Big( \frac{E_0}{T_H} \Big) ^3 \Big(\frac{T_H}{E_{\gamma}}\Big)^{3/2} \;,\; E_{\gamma} < T_H ;\\
  \Big( \frac{E_0}{E_{\gamma}} \Big)^3 , E_{\gamma} \ge T_H,
 \end{array} \right.
  \nonumber
\end{eqnarray}
where $E_0=10^5$, and all energies are measured in GeVs. The
parametrization is valid for $E_\gamma
> m_\pi/2$. The average energy of this distribution is
\begin{eqnarray}
\bar E_{\gamma} \approx 2 T_H \frac {3 -
2\sqrt{\frac{E_{min}}{T_H}}} {4\sqrt{\frac{T_H}{E_{min}}}-3}
\approx 15 \sqrt{T_H} \approx
\nonumber \\
\approx 1300 \Delta t^{-1/6} \; ,
\end{eqnarray}
where we used, just for example, $E_{min}=100 \rm GeV$, and
$\Delta t$ is the remaining PBH lifetime (in seconds), which is
connected with initial Hawking temperature by the simple relation:
\begin{equation}
\Delta t = 4.7 \times 10^{11} \; T_H^{-3} \; .
\end{equation}

\begin{figure}
\begin{center}
\noindent
\includegraphics [trim=1 35 1 10, width=0.45\textwidth]{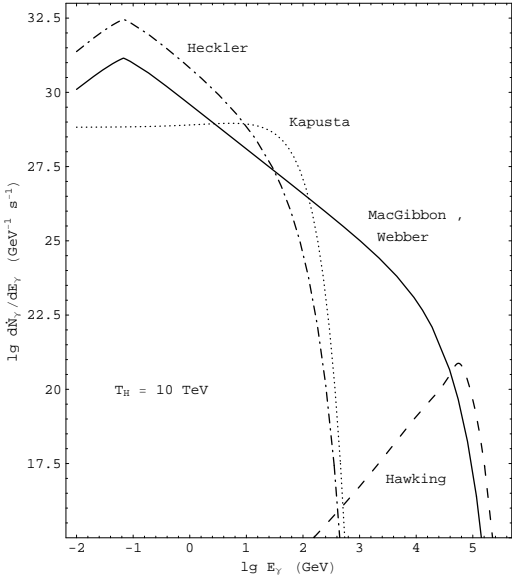}
\end{center}
\caption{Instantaneous photon spectra from a black hole with
Hawking temperature $T_H=10$ TeV.}\label{fig1}
\end{figure}

\begin{figure}[!t]
\begin{center}
\noindent
\includegraphics [trim=1 35 1 1, width=0.45\textwidth]{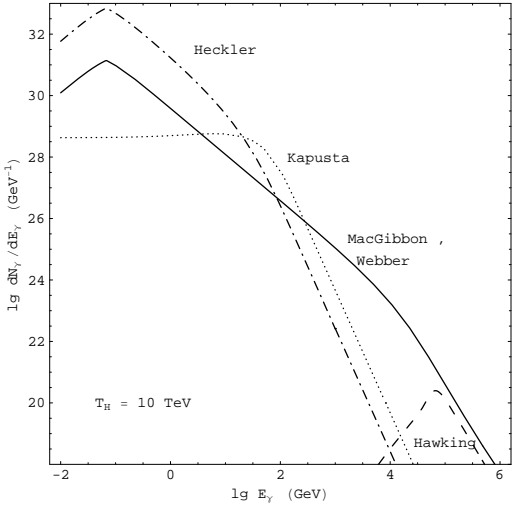}
\end{center}
\caption{Time-integrated photon spectra from a black hole with
initial Hawking temperature $T_H=10$ TeV.}\label{fig2}
\end{figure}

\subsection{Heckler model}

According to Heckler's idea \cite{Heckler:1997jv}, once a black
hole, in course of an evaporation, reaches some critical
temperature $T_{crit}$, the emitted Hawking radiation
begins to interact with itself and forms a nearly thermal
chromosphere. This assumption is based on a simple argument: at
relativistic energies, the cross section for gluon bremsstrahlung
in a $qq$ or $q\bar q$ collision is approximately constant (up to
logarithms of energy) while the density of emitted particles
around the BH is proportional to BH temperature. The evaporation
of BH creates a radiation shell which propagates outward at the
speed of light. The quark and gluon spectrum in the observer rest
frame is obtained by boosting a thermal spectrum at
$T_{crit}=\Lambda_{QCD}$ (and $\Lambda_{QCD}=200\rm MeV$) with the
Lorentz factor $\gamma_{ch} \approx 0.22
(T_{H}/\Lambda_{QCD})^{1/2}$ of the outer surface of the thermal
chromosphere,
\begin{eqnarray}
\label{dNdQH} \frac{d \dot N_j}{dQ} = \sigma_j \frac{\gamma_{ch}^2
r_{ch}^2 Q^2}{2 \pi^2} \times \;\;\;\;\;\;\;\;\;\;\;\;\;\;\;\;\;
 \\ \times  \int \limits_0^1 \frac {(1-\beta \cos\theta) \cos \theta \; d\Omega}
 {\exp [\gamma_{ch} Q (1-\beta \cos\theta)/ T_{ch}] - (- 1)^{2s} }
 \; .  \nonumber
\end{eqnarray}
In this formula the integration over the surface of the
chromospere of radius $r_{ch} = \gamma_{ch}/ \Lambda_{QCD}$ has
been carried out. The observed photon spectrum of the chromosphere
is a convolution of the quark-gluon spectrum given by
(\ref{dNdQH}) with the quark-pion fragmentation function and the
Lorentz-boosted spectrum from $\pi^0$ decay.

The high energy behavior of time-integrated spectrum is
($E_\gamma$ is in GeVs)
\begin{equation}
\frac{dN}{dE_{\gamma}} = 3\times 10^{34} E^{-4}_{\gamma} \; .
\label{dNdEHEH}
\end{equation}
The average energy of time-integrated spectrum is
\begin{equation}
\bar E_\gamma \approx 0.17 \; T_H^{1/4} \approx 1.6 \Delta t  ^
{-1/12}.
\end{equation}

\subsection{Kapusta \& Daghigh model}

This approach \cite{Daghigh:2001gy} differs from those of Heckler
mainly in two respects. It is assumed that the hadronization of
quarks occurs before the thermal freeze-out and the onset of
free-streaming. It is assumed, further, that it happens suddenly
at a temperatute $T_f$ in the range $100-140$ MeV (it is the
parameter of the model), and to this moment all particles whose
mass is greater than $T_f$ have annihilated leaving only photons,
electrons, muons and pions. Photons emitted by BH come from two
sources. Either they are emitted directly from a (boosted)
black-body spectrum or they arise from the $\pi^0$-decay.

The formula for particle spectrum from the outer edge of the
chromosphere is analogous to (\ref{dNdQH}). For the concrete
calculation, we used $T_f=0.12$ GeV. The values of $\gamma_{ch}$
and $r_{ch}$ obtained in \cite{Daghigh:2001gy} are
\begin{equation}
\gamma_{ch} \approx 0.22 \sqrt { \frac {T_H}{T_f} } \; , \; r_{ch}
\approx \frac{0.89}{T_f} \sqrt { \frac {T_H}{T_f} } \; .
\label{Kapusta-g-r}
\end{equation}

The gamma ray flux from the $\pi^0$ - decay in the limit $E_\gamma
\gg m_\pi$ is given by the simple formula \cite{Daghigh:2001gy}
\begin{equation}
\frac{d \dot N^{\pi^0}}{dE_\gamma} = \frac{4 r_{ch}^2
T_f^2}{\pi^2} \sum \limits_{n=1}^{\infty} \frac{\exp({-n E_\gamma
/ 2\gamma_{ch} T_f})}{n^2} ,
\end{equation}
and we used it to estimate photon flux from this channel even at
$E_\gamma \sim m_\pi$.

In this case, the high-energy behavior of time-integrated photon
flux is similar to (\ref{dNdEHEH}), but the flux is approximately
1.5 orders of magnitude larger:
\begin{equation}
\frac{dN}{dE_{\gamma}} = \frac{M_{Pl}^2 T_f} {26 E_\gamma^4} = 6.9
\times 10^{35} \; E_\gamma^{-4}.
\end{equation}
The average energy is given by
\begin{equation}
\bar E_{\gamma} \approx 0.36 \sqrt {T_H} \approx 32 \Delta t
^{-1/6} \; .
\end{equation}

\section{Experimental search of PBH bursts with ground based
detectors}

\begin{figure}
\begin{center}
\noindent
\includegraphics [trim = 10 25 0 0 , width=0.4 \textwidth]{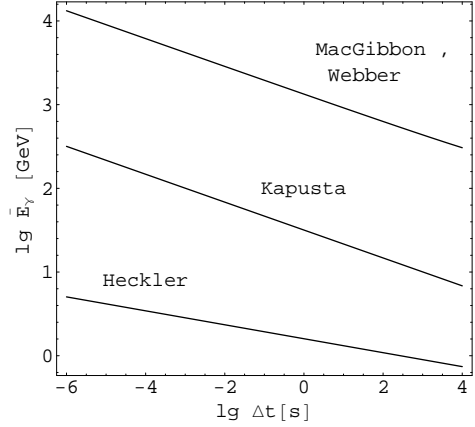}
\end{center}
\caption{Average energy of gamma rays emitted from a PBH, as a
function of left PBH lifetime. Upper curve corresponds to
$E_{min}=100$GeV. }\label{fig3}
\end{figure}

One can see from Fig. \ref{fig3} that the average photon energy as
a function of PBH lifetime is highly model dependent. Therefore,
various methods should be used for the search. Up to now the
search of high energy gamma radiation from PBH with ground-based
arrays had been carried out using the predictions of the model of
\cite{MacGibbon:1990zk}, and, specifically, data on distribution
of events (extensive air showers (EAS)) in time and space were
used. The results were reported in \cite{Alexandreas:1993zx,
Amenomori, Connaughton:1998ac, Funk, Linton:2006yu}, time
intervals used were from 1 to 10 seconds.

\begin{figure}
\begin{center}
\noindent
\includegraphics [trim=0 50 0 0 , width=0.48 \textwidth]{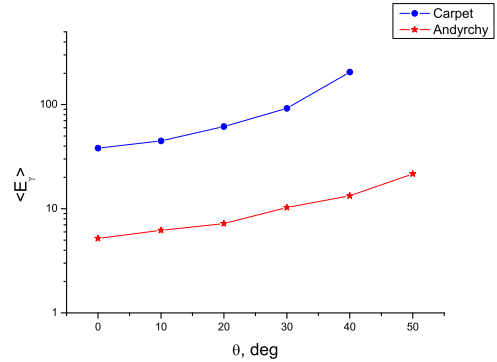}
\end{center}
\caption{Average photon energy vs. zenith angle for model of
\cite{Daghigh:2001gy} .}\label{fig4}
\end{figure}
\begin{figure}
\begin{center}
\noindent
\includegraphics [trim=0 50 0 0 , width=0.45 \textwidth]{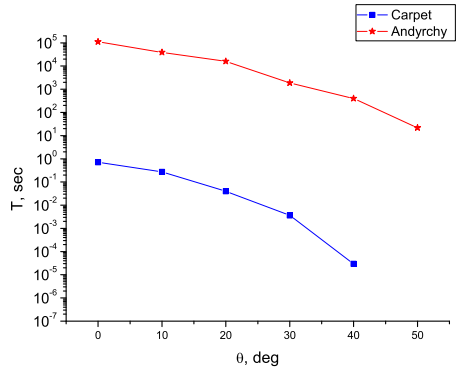}
\end{center}
\caption{Time duration of gamma burst vs. zenith angle for model
of \cite{Daghigh:2001gy}.}\label{fig5}
\end{figure}

If the chromosphere models of black hole evaporation are close to
reality the average photon energy at last instants of PBH life is
rather low. In the model of hadronic chromosphere developed in
\cite{Daghigh:2001gy} this energy is about 30 GeV for $\Delta t
\sim 1 \rm s$. A search of PBH bursts in this case is possible
using the method suggested in works of EAS-TOP collaboration
\cite{Aglietta:1996su} and Baksan group \cite{Petkov, Smirnov}.
Shower particles generated in atmosphere by photons with energy
$\sim 30 \div 100 $ GeV are strongly absorbed before reaching the
detector level, so, the average number of signals in the detector
module is smaller than 1. Therefore, in this energy range PBH
bursts can be sought for by operating with the modules in single
particle mode, that is, by measuring the single particle counting
rate of the individual modules. The primary arrival directions of
photons are not measured, and bursts can be detected only as
transients (short-time increases) of the cosmic ray counting rate.

Using the photon spectra from PBH bursts calculated above and
registration efficiencies of photons (as functions of zenith
angle, the array altitude and features of concrete modules) one
can calculate the average energies of PBH bursts for a given array
and, approximately, corresponding time durations of these bursts.


On Fig. \ref{fig4} the dependencies of average photon energy on
zenith angle in model of \cite{Daghigh:2001gy} for two arrays of
Baksan observatory are shown. Andyrchy and Carpet arrays are
situated practically on the same place (horizontal distance
between them is about 1 km), but their altitudes above sea level
are different: Andyrchy - 2060 m, Carpet - 1700 m. On Fig.
\ref{fig5}, time duration of bursts as a function of zenith angle
is given for the same arrays. One can see that duration of burst
strongly depends on the zenith angle.



In summary, it is shown that predictions for high energy photon
spectra from PBH burst depend rather strongly on the model used
for description of final stage of PBH evaporation. In particular,
chromosphere models predict too steep spectra in TeV region. It is
argued that, if such models are correct, the practical search of
PBH bursts can be carried out using the technique developed by
EAS-TOP and Baksan groups for a search for $\gamma$-bursts at
energies $E>10$ GeV. It is shown that in such search the average
energy and time duration of the signal depend on the zenith angle
of its arrival and on the altitude of the array above sea level.

The work was supported by Russian Foundation for Basic Research
(grant 06-02-16135).

\bibliography{icrc0505}

\begin{thebibliography}{10}

\bibitem{Aglietta:1996su}
M.~Aglietta et~al.
\newblock {Search for gamma ray bursts at photon energies $E \ge 10 \rm GeV$
  and $E \ge 80$ TeV}.
\newblock {\em Astrophys. J.}, 469:305--310, 1996.

\bibitem{Alexandreas:1993zx}
D.~E. Alexandreas et~al.
\newblock New limit on the rate density of evaporating black holes.
\newblock {\em Phys. Rev. Lett.}, 71:2524--2527, 1993.

\bibitem{Amenomori}
M.~Amenomori et~al.
\newblock {Search for 10 TeV Gamma Bursts from Evaporating Primordial Black
  Holes with the Tibet Air Shower Array}.
\newblock In {\em International Cosmic Ray Conference}, volume~2, pages
  112--115, 1995.

\bibitem{Carr:2003bj}
Bernard~J. Carr.
\newblock Primordial black holes as a probe of cosmology and high energy
  physics.
\newblock {\em Lect. Notes Phys.}, 631:301--321, 2003.

\bibitem{Chisholm:2005vm}
James~R. Chisholm.
\newblock Clustering of primordial black holes: Basic results.
\newblock {\em Phys. Rev.}, D73:083504, 2006.

\bibitem{Connaughton:1998ac}
V.~Connaughton et~al.
\newblock A search for tev gamma-ray bursts on a 1-second time- scale.
\newblock {\em Astropart. Phys.}, 8:179--191, 1998.

\bibitem{Daghigh:2001gy}
R.~G. Daghigh and Joseph~I. Kapusta.
\newblock High temperature matter and gamma ray spectra from microscopic black
  holes.
\newblock {\em Phys. Rev.}, D65:064028, 2002.

\bibitem{Funk}
B.~Funk et~al.
\newblock {Search for TeV gamma-rays from evaporating primordial black holes}.
\newblock In {\em International Cosmic Ray Conference}, volume~2, pages
  104--107, 1995.

\bibitem{Heckler:1997jv}
A.~Heckler.
\newblock Calculation of the emergent spectrum and observation of primordial
  black holes.
\newblock {\em Phys. Rev. Lett.}, 78:3430--3433, 1997.

\bibitem{Linton:2006yu}
E.~T. Linton et~al.
\newblock A new search for primordial black hole evaporations using the whipple
  gamma-ray telescope.
\newblock {\em JCAP}, 0601:013, 2006.

\bibitem{MacGibbon:1990zk}
J.~H. MacGibbon and B.~R. Webber.
\newblock Quark and gluon jet emission from primordial black holes: The
  instantaneous spectra.
\newblock {\em Phys. Rev.}, D41:3052--3079, 1990.

\bibitem{Petkov}
V.~B. Petkov et~al.
\newblock {Search for high energy gamma bursts}.
\newblock {\em Kinematics and physics of celestial bodies}, 3:234, 2003.

\bibitem{Smirnov}
D.~V. Smirnov et~al.
\newblock {Search for high energy cosmic gamma bursts at Andyrchy array}.
\newblock {\em Bull. Rus. Acad. Sci., ser. phys.}, 69N3:413, 2005.

\end{thebibliography}

\bibliographystyle{plain}

\end{document}